\documentclass[conference]{IEEEtran}
\IEEEoverridecommandlockouts
\usepackage{cite}
\usepackage{url}
\usepackage{amsmath,amssymb,amsfonts}
\usepackage{algorithm}
\usepackage{algpseudocode}
\usepackage{graphicx}
\usepackage{textcomp}
\usepackage{xcolor}
\usepackage{comment}
\usepackage{authblk}

\def\BibTeX{{\rm B\kern-.05em{\sc i\kern-.025em b}\kern-.08em
    T\kern-.1667em\lower.7ex\hbox{E}\kern-.125emX}}

\newcommand\red[1]{\textcolor{red}{#1}}
\newcommand{\hz}[1]{
{\color{blue} #1 - HZ}
}

\begin{document}

\title{ReVeal: A Physics-Informed Neural Network for High-Fidelity Radio Environment Mapping}

\author{
    Mukaram Shahid, 
    Kunal Das, 
    Hadia Ushaq, 
    Hongwei Zhang, \\
    Jimming Song, 
    Daji Qiao,
    Sarath Babu,
    Yong Guan,
    Zhengyuan Zhu, 
    Arsalan Ahmed
}
\affil{
    Iowa State University 
}

\maketitle

\begin{abstract}

Accurately mapping the radio environment (e.g., identifying wireless signal strength at specific frequency bands and geographic locations) is crucial for efficient spectrum sharing, enabling secondary users (SUs) to access underutilized spectrum bands while protecting primary users (PUs). 
    However, current models are either not generalizable due to 
    shadowing, interference, and fading or are computationally too expensive, limiting real-world applicability. 
To address the shortcomings of existing models, we derive a second-order partial differential equation (PDE) for the Received Signal Strength Indicator (RSSI) based on a statistical model used in the literature. 
    We then propose ReVeal (Re-constructor and Visualizer of Spectrum Landscape), a novel Physics-Informed Neural Network (PINN) that integrates the PDE residual into a neural network loss function to accurately model the radio environment 
    based on sparse RF sensor measurements. 
ReVeal is validated using real-world measurement data from the rural and suburban areas of the ARA testbed and benchmarked against existing methods. 
ReVeal outperforms the existing methods 
in predicting the radio environment; for instance, with a root mean square error (RMSE) of only 1.95 dB, ReVeal achieves an accuracy that is an order of magnitude higher than existing methods such as the 3GPP and ITU-R channel models, ray-tracing, and neural networks. 
ReVeal achieves both high accuracy and low computational complexity while only requiring sparse RF sampling, for instance, only requiring 30 training sample points across an area of 514 square kilometers. 
    These promising results demonstrate ReVeal's 
    potential to help advance the state of the art in spectrum management by enabling precise interference management between PUs and SUs. 

\end{abstract}


\section{Introduction}

Existing spectrum sharing frameworks, such as those implemented in the TV White Space (TVWS) database and CBRS Spectrum Access System (SAS), heavily rely on conventional statistical models. These models struggle to accurately capture the real-world spectrum occupancy and do not generalize well enough to capture shadowing and fading caused by different kinds of terrain and environmental conditions. This leads to conservative approaches that over-protect the primary users~(PUs) and create discrepancies in channel availability for spectrum re-use \cite{Magazine,DB_Critique,DB_Implementation}. 
    In the meantime, deterministic models such as ray tracing 
require precise characterization of the complete propagation environment such as 
vegetation, trees, buildings,  and material properties. Any errors in accurately defining these site-specific characteristics can degrade the models' accuracy. In addition, these deterministic models are usually too computationally expensive to be useful for at-scale, online spectrum management in dynamic radio environments. 
    The existing stochastic and deterministic models also typically require the transmitter's operational parameters, such as effective isotropic radiated power (EIRP), transmitter location, and antenna characteristics, which may well be unavailable in real-world scenarios (e.g., where strong privacy or military secrecy are desired). 
The aforementioned drawbacks call for new models that are generically applicable to diverse environments and that are highly accurate in capturing the impact of transmitters and environmental factors (e.g., vegetation, trees, buildings) on receiver signal strength while not requiring comprehensive, highly-accurate information about the transmitters and environment. 

To address the above challenge, data-driven modeling via Spectrum Cartography (SC) 
offers a promising solution avenue. 
In SC, ground-truth wireless signal measurements from sparsely-distributed RF  sensors are used to accurately generate the radio environment map (REM) in the geographical area of interest 
\cite{Magazine,Bhattarai2018,vtechworks_2024,Spectrum_carography,cartography_techniques}. 
    In particular, SC treats radio environment mapping as an ill-posed inverse problem, where the locations and RF parameters of the transmitters are not available, and SC uses the spatial relationship between the measurements to regenerate the REMs \cite{Subash,SC_survey,NTIA_spectrum_cartography}. The generated REMs have a wide range of applications in wireless communications, for instance, dynamically identifying white spaces for efficient spectrum sharing, optimizing power control for interference management, and facilitating seamless handover \cite{cartography_techniques,Spectrum_carography}. 

Despite their promises, existing methods for generating REMs face significant limitations. For example, techniques such as Kriging and Tensor Decomposition assume a uniform spatial structure, failing to capture complex variations of signal strength often encountered in real-world scenarios. They also typically require dense data, introducing significant 
computational and sensing costs \cite{Subash}, \cite{Block_tensor_decomposition}. Similarly, deep learning approaches, while being powerful in matrix or tensor completion tasks, often lack interpretability and require vast labeled datasets, which are impractical to collect in real-world scenarios.

To fill the gap in radio environment mapping, we propose ReVeal (Re-constructor and Visualizer of Spectrum Landscape), a novel Physics-Informed Neural Network (PINN) architecture for blind spectrum cartography. 
ReVeal uses a partial differential equation (PDE) to characterize the spatial variations of wireless signal strength, and then it incorporates the PDE as a physical constraint to a Fully Connected Neural Network (FCNN) with random dropouts. 
This innovative approach enables ReVeal to achieve high accuracy with minimal data, capturing real-world signal variations without requiring prior knowledge of transmitter parameters or detailed environmental information (e.g., terrain). The key contributions of this paper are as follows:
\begin{itemize}
    \item \textbf{Introduction of PINN to  Spectrum Cartography:} This paper pioneers the use of PINNs in the domain of blind spectrum cartography, offering a new way of integrating physical laws with data-driven learning to improve the accuracy and efficiency of radio environment mapping 
    without knowing the operational parameters of the transmitters nor environments.

    \item \textbf{PDE Form of Path Loss Model:} The paper derives a novel partial differential equation 
     based on a well-known statistical path loss model, enabling ReVeal to model spatial variations in signal strength caused by shadowing. This enables the model to accurately capture the impact of shadowing and other environmental factors without any prior knowledge of the 
     environment. The PDE-based loss shows its edge in capturing the shadowing distribution, which is not possible when simple empirical path loss models are used with the PINN.

    \item \textbf{Data- and Computation-Efficient Solution:}
    ReVeal requires significantly fewer sample points compared to traditional techniques, achieving high accuracy with minimal data. This data efficiency, combined with the model's fast convergence and optimized architecture, makes ReVeal suitable for at-scale, online spectrum management in dynamic radio environments. 

    \item \textbf{Real-world Outdoor Evaluation:}
    ReVeal has been evaluated using the ARA \cite{ARA_design_implementation} testbed under varying rural and suburban terrains and channel conditions. We also benchmark the performance of ReVeal with other statistical, deterministic, geospatial, and neural network models using real-world data collected over an area of \mbox{19.4\,km $\times$ 26.5\,km} in ARA. The datasets and software implementation of ReVeal will be publicly released for community access after the paper review period.

\end{itemize}

The rest of the paper is structured as follows:  Section~\ref{sec:related_work} discusses 
related work, 
Section~\ref{sec:methodology} presents ReVeal, 
Section~\ref{sec:experimental_setup} presents the experiment evaluation plan, 
Section~\ref{sec:results} presents the experimental results, 
and Section~\ref{sec:conclusion} makes the concluding remarks. 

\section{Related Work }

\label{sec:related_work}

In what follows, we first review the stochastic and deterministic models typically used in today's spectrum management practice, then we discuss geospatial models, deep learning models, as well as physics-informed deep learning.

\paragraph{Stochastic and Deterministic Models}

Channel modeling has been an integral part of wireless communication systems design, 
providing information about signal propagation,  which in turn can be used for interference management and network optimization~\cite{Channel_measurement_survey}. Stochastic modeling techniques revolve around the use of statistical distributions or empirical equations to determine the distribution of signals given the operational parameters of the transmitter (e.g., location, height and azimuth of the antenna), and the information about the line-of-sight between the transmitter and the receiver. 
The use of probabilistic techniques and statistical distribution in these models reduces the computational complexity. 
however, such simplicity often comes at the expense of accuracy~\cite{Tataria2020}. Focusing on summary statistical distributions instead of specific instances and without capturing the specific environment features (e.g., vegetation, trees, and buildings) at a given site, these models cannot precisely characterize the site-specific wireless channel behavior due to significant spatial variation in wireless signal propagation shadowing. 

There exist deterministic models such as ray tracing that are site specific and enable the precise modeling of the propagation environment considering the geographical scene, properties of the material, and the scatters between the transmitter and receiver~\cite{ray_tracing_based_model,Nueral_Ray_Tracing}. By considering principles of physics such as reflection, defraction, and scattering, deterministic models can precisely calculate the path loss, delay, and angle of each reflected component reaching the receiver. However, precision comes at the expense of computation and space complexity. In addition, in real-world situations where it is difficult to precisely characterize the environment, such deterministic models can have large errors too as we will demonstrate in Section~\ref{sec:results}. 

Both stochastic and deterministic models require prior operational information about the transmitter, such as height, azimuth, and EIRP, which may not be realistic in Radio Dynamic Zones~(RDZs) where multiple users utilize spectrum as a shared common resource pool. Moreover, these models are not capable of integrating real-time spectrum usage data from 
RF sensors deployed in RDZs~\cite{RDZ}.

\paragraph{Geospatial 
and Deep Learning Models}

Geospatial interpolation techniques have been the center of attention in the wireless community for generating REMs. Approaches such as Kriging~\cite{Kriging} and inverse weighted distance have been of use in modeling the spectrum occupancy based on the sparsely collected data points. However, these techniques work with the assumption of spatial stationarity and struggle to capture the nonlinear relationships, often experienced in modeling wireless channels, due to the presence of shadowing and wireless interference~\cite{Kriging}. Furthermore, these models generally lack the ability to accommodate 
new terrains and varying spatial resolutions. 

Deep Learning~(DL) algorithms, on the other hand, are able to learn complex nonlinear spatial relationships from sparse training data \cite{DeepREM,ProSpire}. Various deep learning~(DL) models such as CNN~\cite{CNN}, U-NET~\cite{U_NET}, and GANs~\cite{GAN} have been proposed to generate spatio-temporal spectrum maps. However, DL approaches require significant amount of training data, and collecting data from real-world deployments is a time-consuming task. Furthermore, achieving a fixed dense deployment of RF sensors is often impractical due to cost and data collection overhead.

\paragraph{Physics-Informed Deep Learning}

Physics-Informed Deep Learning (PIDL) has emerged as a new compelling method to solve PDEs for both forward and inverse problems. Finite Element Methods (FEMs) have been the key in solving PDEs in different engineering problems. However, while solving PDEs, FEMs are not capable of integrating real-world data without complex computationally expensive data assimilation techniques \cite{PINN_to_PIKAN}. This limitation prevents FEMs from fully utilizing measurement data, which can cause valuable system insights to be overlooked \cite{Possion_FEM}. 
    In contrast, neural networks are naturally suited for data assimilation, as they can be trained using data of varying fidelity and modality. Physics-Informed Neural Networks (PINNs) have been developed to bridge the gap between data-driven and physics-based methods, especially in cases where partial knowledge of the physical laws and sparse measurement data are available. By embedding physical laws directly into the neural network through 
    residual loss terms in the objective function, PINNs can enforce the governing PDEs as soft constraints, which enable PINNs to solve forward and inverse problems using sparse and noisy data \cite{PINN_RAISSI}. 

PINNs so far have been widely used in applications ranging from acoustic engineering to the modeling of flow dynamics and modeling of electromagnetic fields but have not been fully explored in the areas of wireless channel modeling and 
spectrum sharing. In this work, we will explore and develop an innovative architecture that can leverage a PINN model 
for spectrum cartography in the TVWS band utilizing the real-world data from the ARA testbed.

\section{ReVeal: PINN for Radio Environment Mapping}
\label{sec:methodology}

\begin{figure}[!htbp]
        \centering
        \includegraphics[width=1\columnwidth]{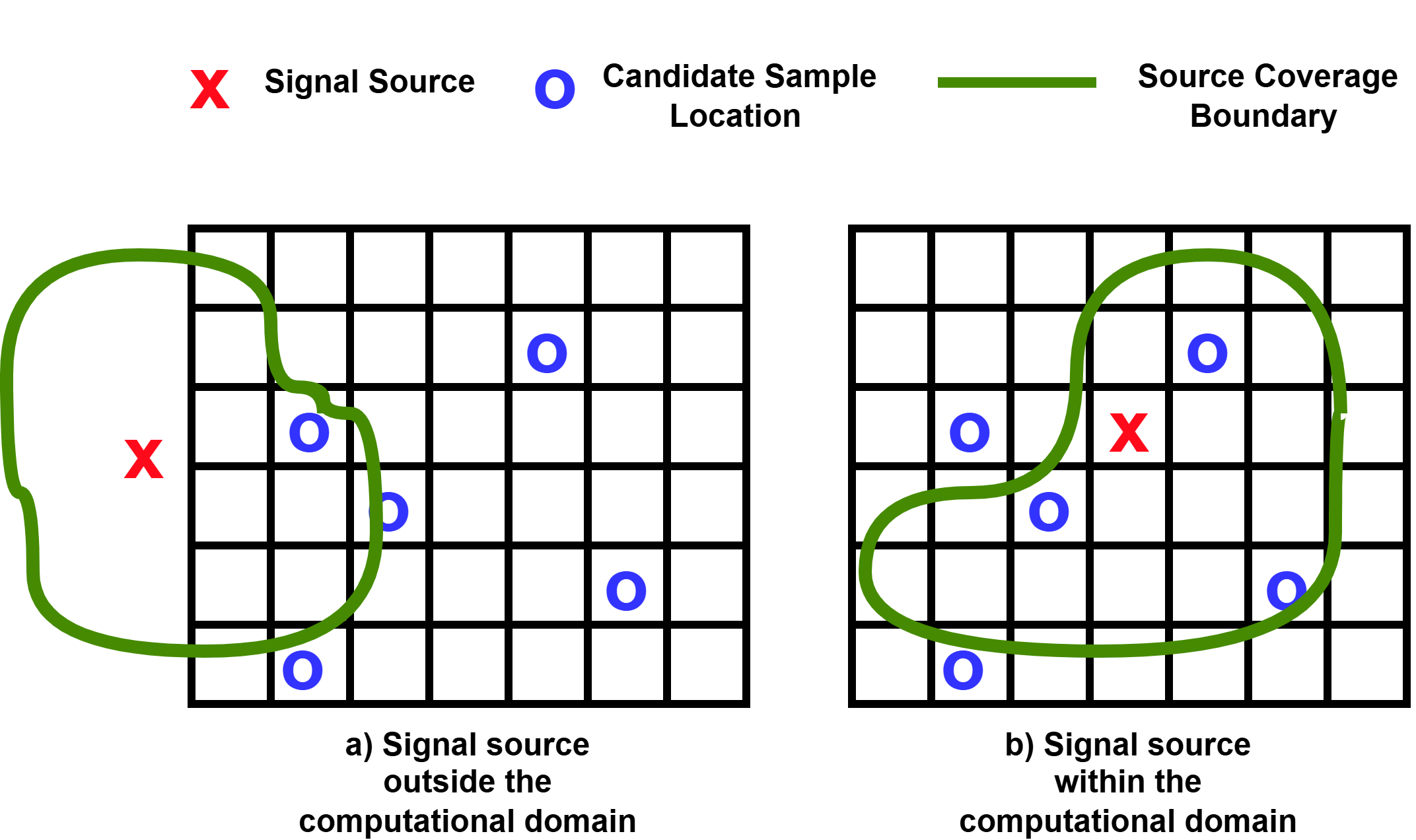}
        \caption{Visualization of signal source placement relative to the domain/geographic-region of interest. The signal source~(denoted as $X$) influences the candidate sample locations~(denoted as O) differently depending on its position, altering the source coverage boundary (denoted in green color).}
        \label{fig: Methodology}
\end{figure}

Here we focus on developing the spatial radio environment map~(REM) of a specific geographical region of interest denoted as domain~\(\mathbb{D}\). The domain~\(\mathbb{D}\) is discretized into $I \times J$  
equally sized cells, each cell 
representing a spatial location within the region of interest. 
Assuming there is a transmitter~$X$ which could be within or outside~\(\mathbb{D}\), as shown in Figure~\ref{fig: Methodology}. 
\begin{table}[!b]
\centering
\caption{Notation Summary}
\label{tab:notation}
\small 
\begin{tabular}{|p{1.5cm}| p{6cm}|}\hline
\renewcommand{\arraystretch}{1.4}
\textbf{Notation} & \textbf{Description} \\ \hline\hline
$\mathbb{D}$ & Domain representing the specific geographical region of interest \\ \hline
$I \times J $ 
& Grid dimensions of the domain $\mathbb{D}$ \\ \hline
$\Omega_n$ & Set of RF sensors sparsely deployed across the region $\mathbb{D}$; $n = 1 \ldots N$ \\ \hline
$P^{\text{obs}}(\Omega_n, C)$ & Observed power measurements (RSSI) at sensor locations $\Omega_n$ for channel $C$ \\ \hline
$\mathbb{C}$ & Set of channels or spectrum of interest, partitioned into discrete frequency bands \\ \hline
$P^{\text{pred}}$ & Predicted power measurements (RSSI) by the model at sensor locations \\ \hline
$L$ & Objective function representing the error between observed and predicted values \\ \hline
\end{tabular}
\end{table}
The transmitter's location and other parameters (e.g., transmission power, antenna) 
are unknown. 

A set of RF sensors $\Omega_n$ ($n = 1 \ldots N$) 
are sparsely deployed across $\mathbb{D}$ in a randomly manner to observe the Received Signal Strength Indicator (RSSI) at selected locations. 
Each sensor provides observations of RSSI or power measurements $P^{\text{obs}}$~$(\Omega_n, C)$ for each channel $C$ from a set of communication channels $\mathbb{C}$. 
The set of channels $\mathbb{C}$ represents the spectrum of interest, partitioned into discrete frequency bands, with each band corresponding to a unique channel. The measurements collected by these sensors are sparse and are impacted by shadowing and large-scale path-loss, which vary across both spatial and spectral dimensions. 
Our main objective in spectrum cartography (SC) 
is to model a function that, given any location in $\mathbb{D}$, generates an accurate prediction of the RSSI at the location, thus enabling the generation of the radio environment map for domain $\mathbb{D}$. Mathematically, the objective is to minimize  the error~($L$) between the expected observed RSSI~($P^{\text{obs}}$) and expected model-predicted RSSI~($P^{\text{pred}}$), as shown below: 
\begin{equation}
    L = \sum_{n=1}^{N} | E[{P}^{\text{pred}}(\Omega_n, C)] - E[P^{\text{obs}}(\Omega_n, C)] |^2. 
    \label{eq:objective}
\end{equation}


\subsection{Physics-Informed Neural Network in ReVeal}

Physics-Informed Neural Networks (PINNs) are recent developments in the field of scientific machine learning that utilizes the ability of neural networks to learn the underlying physics.
The idea behind PINNs is to incorporate, as a component of the neural network loss function, the equation governing the underlying physical law
during the training process, where the equation is typically a partial-differential-equation (PDE). The mean squared residual of the governing PDE along with the data-driven loss function is minimized to train the 
neural network. 

\begin{figure}[!htbp]
        \centering
        \includegraphics[width=1\columnwidth]{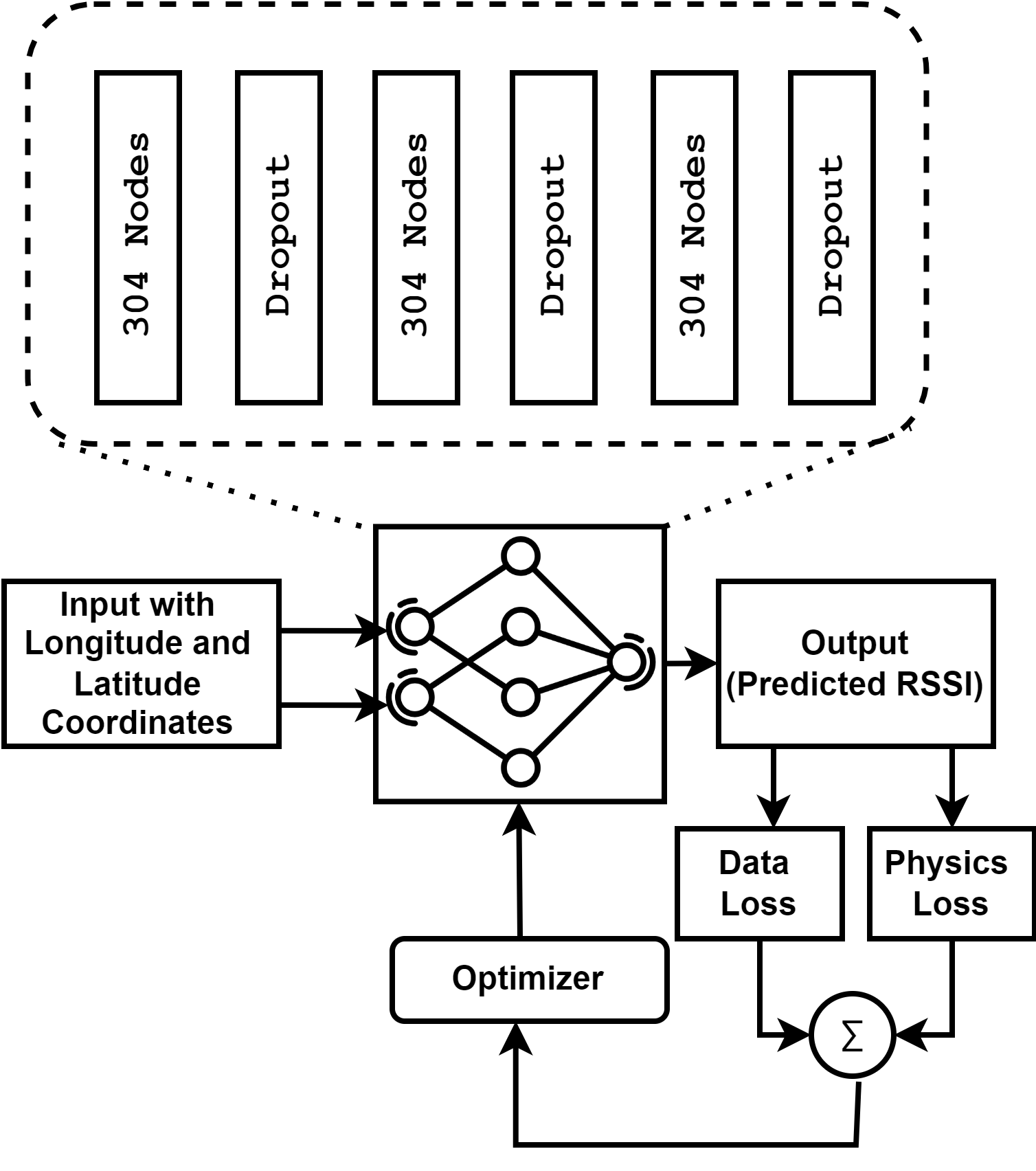}
        \caption{PINN architecture of ReVeal}
        \label{fig: Pinn_architecture}
\end{figure}

To solve the spectrum cartography (SC) problem, 
we propose a PINN architecture as illustrated in Figure\ref{fig: Pinn_architecture}.
The input for the network is the set of spatial locations of the measurement points $\Omega_n$ representing the geographical locations 
of the RF sensors within the domain $\mathbb{D}$. For a channel~$C$, the output of the PINN $P^{\text{pred}}$ represents the expected RSSIs at the locations given to the PINN. 
During training, $P^{\text{pred}}$ is compared with the observed values at each training sampling 
point to compute the data-driven loss. Apart from the data-driven loss term, the residual of the governing-physics PDE is used to help reduce the overall error of the PINN. The calculated loss is sent to the optimizer and is used to update 
the weights and biases of the neural network. 
To overcome over-fitting, multiple dropout layers are used in the neural network.


The overall loss function of the PINN forces the model to not only reduce the Mean Square Error~(MSE) based on the measured data but also conform to the physics, helping the model to generalize better during the training process. The overall loss function for the PINN optimization can be defined as

\begin{equation}
L_{\text{total}} = (1- \lambda) L_d + \lambda L_p,
\label{eq:total}
\end{equation}
where, 
\begin{equation}
L_d =  \sum_{i=1}^{N} \left| {E[P^{\text{pred}}}(\Omega_n, C)] - E[{P^{\text{obs}}}(\Omega_n, C)] \right|^2,
\label{eq:Ld}
\end{equation}
and,
\begin{equation}
L_p =  \sum_{i=1}^{N} \left| \nabla^2 {E[P^{\text{pred}}}(\Omega_n, C)]- \nabla^2 E[P^{\text{obs}}(\Omega_n, C)] \right|^2.
\label{eq:Lp}
\end{equation}

\noindent Here, $L_d$ refers to the loss calculated from the labeled data points collected from the sensors. The physics-driven loss $L_p$ is defined by by Eqn.~\eqref{eq:Lp} using the second-order derivative of the predicted RSSI and observed RSSI with respect to the spatial coordinates (see Eqn.~\eqref{eq:sum}). The parameter~$\lambda$ in Eqn.~\eqref{eq:total} is a variable used to tune the weightage given to the data-driven or physics-driven loss during the training of the NN.

\subsection{Governing-Physics PDE in ReVeal}

A key design decision of ReVeal is to choose the physics model governing the spatial dynamics of RSSI. To this end, we derive a second-order PDE based on well-established wireless signal path loss models. 
    Specically, a well-known path loss model is shown by Equation~\eqref{eqn:path_loss} below:
\begin{equation}
P_r(x, y) = P_T  -  \bigg[10\eta \log_{10} \bigg( 
\frac{\sqrt{(x - x_T)^2 + (y - y_T)^2}}{d_0} \bigg)\bigg] + Z_\sigma.
\label{eqn:path_loss}
\end{equation}
It describes the received signal power \( P_r(x, y) \) at any point \( (x, y) \), as a function of the transmit power \( P_T \), distance from the transmitter \( \sqrt{(x - x_T)^2 + (y - y_T)^2} \), path loss exponent $\eta$, reference distance $d_0$, and an added shadowing factor \( Z_\sigma \) representing random spatial variations\footnote{The temporal variation caused by fading is beyond the scope of this study and treated as a part of the future work.}. 
Taking the expectation of both sides of Eqn.~\eqref{eqn:path_loss}, we get

\begin{equation}
\begin{split}
E[P_r(x, y)] = P_T - \bigg[10\eta \log_{10} \bigg( 
\frac{\sqrt{(x - x_T)^2 + (y - y_T)^2}}{d_0} \bigg)\bigg] \\
+ E[Z_\sigma].
\end{split}
\label{eqn:expectation}
\end{equation}


\noindent The second-order partial derivative of Eqn.~\eqref{eqn:expectation} with respect to $x$ results in

\begin{equation}
\frac{\partial^2 E[P_r(x, y)]}{\partial x^2} = \frac{10\eta}{\ln(10)} \cdot \frac{(y - y_T)^2 - (x - x_T)^2}{[(x - x_T)^2 + (y - y_T)^2]^2}  + \frac{\partial^2 E[Z_\sigma]}{\partial x^2}.
\label{eq:d_dx}
\end{equation}
Similarly, the second order partial derivative of Eqn.~\eqref{eqn:path_loss} with respect to \( y \) results in
\begin{equation}
\frac{\partial^2 E[P_r(x, y)]}{\partial y^2} = \frac{10\eta}{\ln(10)} \cdot \frac{(x - x_T)^2 - (y - y_T)^2}{[(x - x_T)^2 + (y - y_T)^2]^2} + \frac{\partial^2 E[Z_\sigma]}{\partial y^2}.
\label{eq:d_dy}
\end{equation}
Adding Eqns.~\eqref{eq:d_dx} and~\eqref{eq:d_dy}, we get

\begin{equation}
\frac{\partial^2 E[P_r(x, y)]}{\partial x^2} + \frac{\partial^2 E[P_r(x, y)]}{\partial y^2} = \frac{\partial^2 E[Z_\sigma]}{\partial x^2}+\frac{\partial^2 E[Z_\sigma]}{\partial y^2}.
\label{eq:sum}
\end{equation}

\noindent The right-hand side of Eqn.~\eqref{eq:sum} denotes the variation in signal strength due to shadowing and any other sources present in the domain. 
    In simple free-space settings, the signal variation due to shadowing is zero, thus the right-hand side of Eqn.~\eqref{eq:sum} is always zero and does not vary across space. 
Since there exists shadowing in real-world settings, the right-hand side of Eqn.~\eqref{eq:sum} becomes non-zero. Thus a good model needs to precisely capture the impact of shadowing and needs to make sure that the second-order derivatives of the predicted signal strength in domain $\mathbb{D}$ as close as possible to the second-order derivatives of the observed signal strength in $\mathbb{D}$. Therefore, we define the physics-driven loss term $L_p$ as shown in Equation~\eqref{eq:Lp}.  


\subsection{Learning Criteria and Algorithm of ReVeal}

During training, the network optimizes the composite loss function that integrates both data-driven and physics-driven losses. After the forward pass, the model predicts the expected RSSI at the training sample 
points. Further, the model calculates the error between the predicted and the observed values using the data-driven loss defined in Equation~\eqref{eq:Ld}. Additionally, leveraging the network's automatic differentiation capability, the model computes the second-order spatial derivatives of the predicted RSSI values. These derivatives are used to evaluate the residuals of the governing-physics PDE, defined in Equation~\eqref{eq:Lp}. The physics-driven loss term minimizes these residuals, enforcing consistency with the governing-physics PDE with respect to the observed RSSI. 
During back-propagation, the optimizer adjusts the network’s weights and biases to minimize the composite loss ensuring both the data-driven and the physics-driven loss terms are considered appropriately. 

The balance between the physics-driven loss term and the data-driven loss term is achieved using the parameter~$\lambda$, which can adjust the relative importance of the two loss components. 
Without the data-driven loss term, the model lacks a starting point for optimization since the physics-driven loss term, based on a PDE, does not provide sufficient guidance for aligning predictions with real-world observations. Therefore, a suitable choice of~$\lambda$ is essential to maintain a right combination of empirical accuracy and physical consistency during the training process, as we will elaborate more in Section~\ref{sec:results}.

\subsection{Optimizing the PINN Architecture}

Hyper-parameter tuning is a crucial yet tedious task in designing machine learning algorithms, including PINNs. Optimal selection of parameters such as number of hidden layers, number of neurons per layer, activation function and learning rate highly impact the performance as well as the convergence of the model. In literature advanced hypermeter tuning algorithms that uses techniques such as grid search and random search for parameter selection such as Autotune \cite{Autotune} and SMAC \cite{SMAC} have been used. However most of these libraries need a predefined search space by the user to find the best parameters against minimizing an objective function. Libraries such as Optuna~\cite{optuna} on the other hand gives users the flexibility to define a dynamic search space and employ advanced optimization techniques such as TPE for dynamic and efficient hyper-parameter tuning without a pre-define grid search for efficient and optimal parameter selection. For this study, we use Optuna~\cite{optuna} as a hyper-parameter optimization library to choose parameters based on the spatially sampled data points, and the resulting hyper-parameters are shown in Table~\ref{tab:hyperparameters}. 


\begin{table}[ht]
\centering
\caption{Hyper-parameters of ReVeal}
\label{tab:hyperparameters}
\begin{tabular}{|l|l|}
\hline
\textbf{Hyper-parameter}         & \textbf{Value}               \\ \hline\hline
Number of input features         & 2                            \\ \hline
Number of hidden layers          & 3                            \\ \hline
Number of neurons per layer      & 304                          \\ \hline
Activation function              & ReLU                         \\ \hline
Dropout rate                     & 0.2                          \\ \hline
Number of output features        & 1                            \\ \hline

Learning rate                    & 0.00369                               \\ \hline

\end{tabular}
\end{table}


Putting all the aforementioned designs together, Algorithm~\ref{alg:PINN-DPZ} summarizes the ReVeal algorithm. 

\begin{algorithm}[!h]
\caption{\textbf{ReVeal Algorithm for Generating Radio Environment Map (REM)}}
\label{alg:PINN-DPZ}
\begin{algorithmic}[1]
    \State Define the domain $\mathbb{D}$ 
    of interest as a grid of $I \times J$ cells.

    \State Define the locations   $\Omega_n$ of the RF sensors within $\mathbb{D}$.

    \State Load the observed RSSI values $P^{\text{obs}}$ from the sensors at locations $\Omega_n$ for each channel $C$.

    \Statex \textbf{[Initialize the PINN model]} 
    \State $model = \text{PINN}(\text{hyper-parameters})$

    \Statex \textbf{[Define the Loss function]}
    \State $L_d 
    = \text{MSE}(E[P^{\text{pred}}(\Omega_n, C)], E[P^{\text{obs}}(\Omega_n, C)])$
    \State $L_p 
    = \text{MSE} \left( \nabla^2 E[P^{\text{pred}}(\Omega_n, C)], \nabla^2 E[P^{\text{obs}}(\Omega_n, C)] \right)$

    \State $L_{\text{total}} 
    = (1-\lambda) L_d + \lambda L_p$
    \Statex \textbf{[Train the PINN model]}    
    \For{epoch = 1 \textbf{to} num\_epochs}
        \For{batch \textbf{in} batches($\Omega_n$, $P^{\text{obs}}$)}
            \State Predict the power measurements $P^{\text{pred}}$ for the  \newline  current batch: $P^{\text{pred}} = model(batch)$
            \State Calculate the total loss: $loss = L_{\text{total}} 
            $
            \State Update the model parameters using the optimizer: \newline $optimizer.step(loss)$
        \EndFor
    \EndFor
    \Statex \textbf{[Obtain the predicted REM]} 
    \State $\text{REM} = model(\mathbb{D})$
    \end{algorithmic}
\end{algorithm}

\section{Experimental Setup} 
\label{sec:experimental_setup}

\begin{figure}[!htbp]
        \centering
        \includegraphics[width=1\columnwidth]{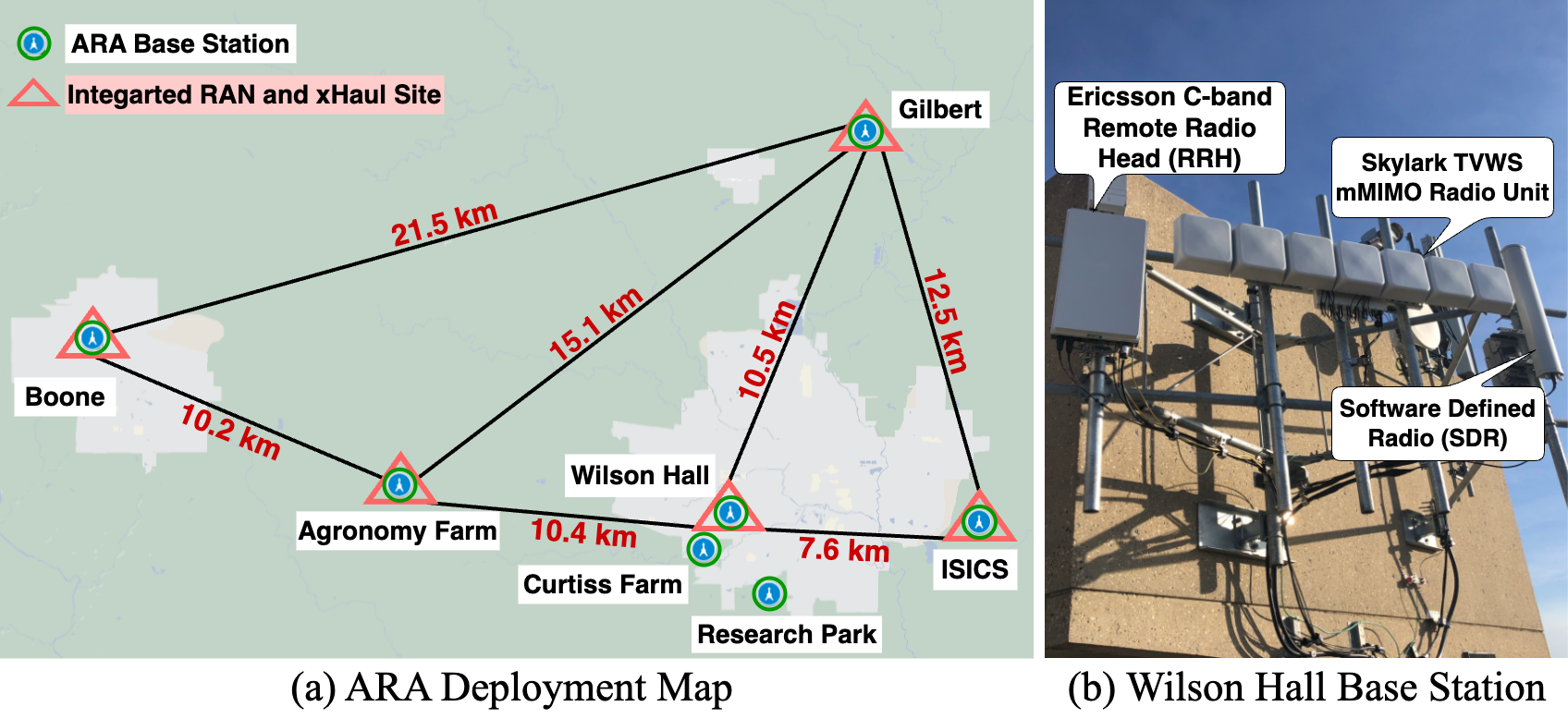}
        \caption{ARA Deployment 
        }
        \label{fig: ARA_deployment}
\end{figure}

We evaluate ReVeal using real-world data collected from the ARA testbed \cite{ARA_design_implementation}. ARA is a first-of-its-kind wireless living lab located around Iowa State University spanning an area over 500 square kilometers, covering research and producer farms along with rural communities of Central Iowa. Figure~\ref{fig: ARA_deployment} 
shows the deployment of seven ARA base station~(BS) sites. Among the seven BSes, Wilson Hall, Boone, ISICS, and Gilbert are equipped with the SkyLark massive MIMO~(mMIMO) system operating in the TVWS band. 
On the other hand, Wilson Hall, Agronomy Farm, Research Park, and Curtis Farm sites includes a C-Band commercial-off-the-shelf~(COTS) mMIMO system from Ericsson operating at 3450--3550\,MHz band. In addition, all BS sites include software-defined-radio~(SDR) operating at 3400--3600\,Mhz supporting end-to-end whole-stack 5G experiments using open-source systems such as OpenAirInterface and srsRAN.

\paragraph{Data Collection Site}

\begin{figure}[!htbp]
        \centering
        \includegraphics[width=.6\columnwidth]{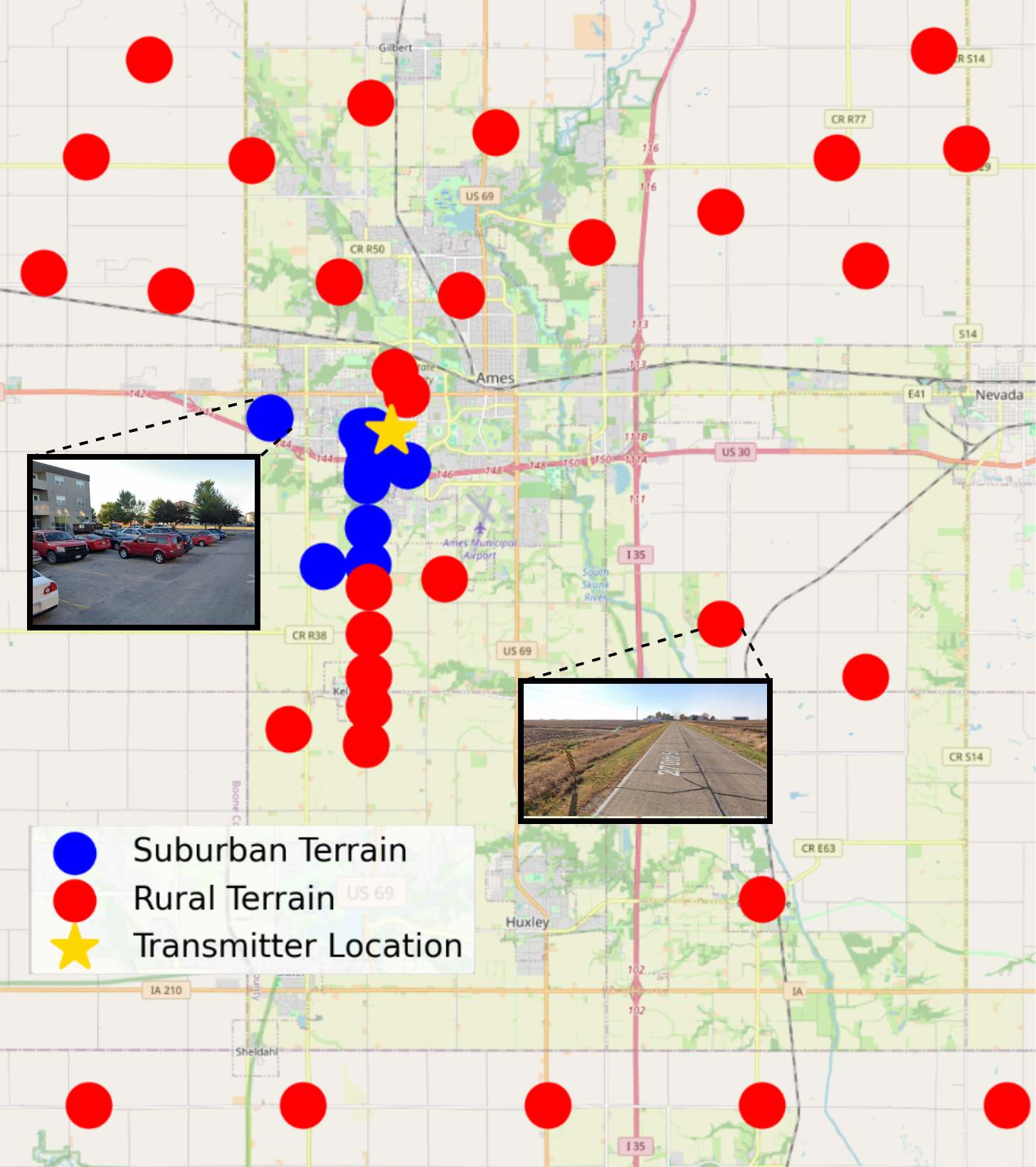}
        \caption{Sampling Locations and Corresponding Terrain Conditions 
        }
        \label{fig:dataSamples}
\end{figure}

\begin{figure}[!htbp]
        \centering
        \includegraphics[width=1\columnwidth]{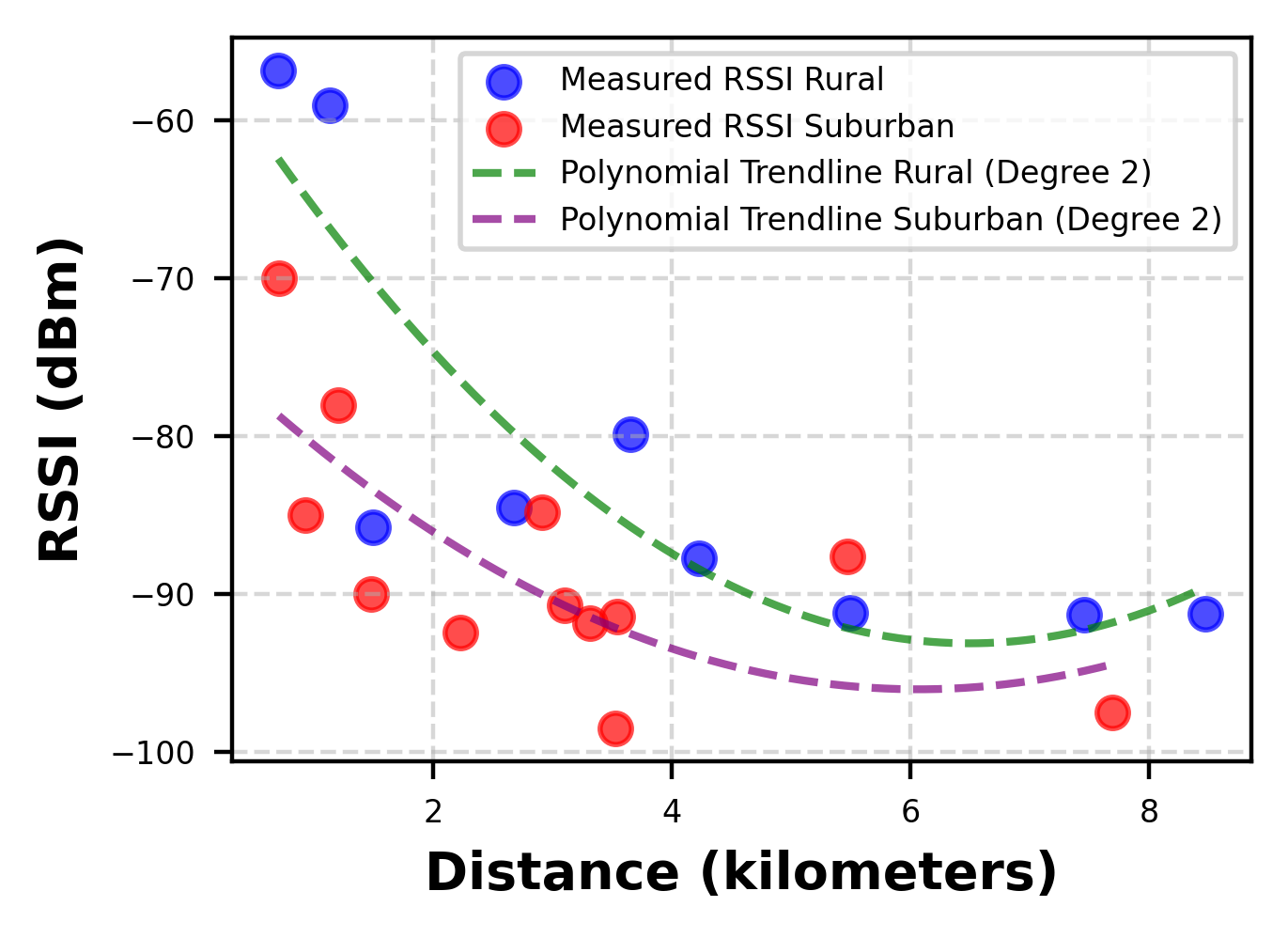}
        \caption{Variation in Channel Conditions across Rural and Suburban Terrains 
        }
        \label{fig:ruralSurburbanChannels}
\end{figure}

For the REM modeling, we collected real-world data using the 
Skylark mMIMO systems deployed at the Wilson Hall base station. The sample points 
were selected 
around the Wilson Hall BS site in an area spanning 514 square kilometers. 
All the other BSes at the time of sampling were turned off, and, before the beginning of the actual sampling process, it was made sure there were no other transmitters operating at the same frequency as the TVWS band used by the Skylark BS. 
The coverage range of the Skylark BS 
around the Wilson Hall site includes diverse 
terrains from rural communities to suburban areas of Downtown Ames that can be seen in Figure~\ref{fig:dataSamples}. These different terrains come with 
different distributions of shadowing and fading effects. 
As shown in Figure~\ref{fig:ruralSurburbanChannels}, the measurement sampling instances show varying levels of shadowing, even when the distance from the BS is the same.


\paragraph{Spectrum Sensing Equipment}

The ARA BS sites have spectrum sensing equipment, for instance, the Keysight N6841-A RF Sensors 
connected to the Keysight N6850A omnidirectional antennas, monitoring spectrum activities across the bands of interest. Apart from the fixed spectrum sensors, the Keysight N9952A FieldFox equipped with an N6850A Omnidirectional antenna and the Keysight NEMO handheld were used to capture the RSSI values at different spatial points of interest around the BS site too.

\paragraph{Spatial Sampling Strategy}

Selecting suitable spatial samples 
is important to reduce the sampling overhead while ensuring necessary modeling accuracy. Spatially balanced sampling is essential in spatial modeling to minimize prediction errors across the whole domain. The primary goal of this approach is to find the most informative sample points out of all the candidate sample locations in the dataset \cite{spatial_sampling}. 
Since the spatial points that are very close to one another are likely to have similar or identical RSSIs, it is desirable to select spatial points that are farther apart and that are as representative as possible of the whole population under consideration. To this end, the Local Pivotal Method (LPM) \cite{grafstra2012spatially} is an effective method, and we use it in deciding the spatial sampling locations in our study (as shown in Figure~\ref{fig:dataSamples}). 

\begin{table*}
\centering
\caption{Performance of Different Path Loss Models Across the Whole Domain
}
\label{tab: metrics}
\normalsize 
\begin{tabular}{|p{6cm}|c|c|c|c|}\hline
\renewcommand{\arraystretch}{1.4}
\textbf{Model} & \textbf{RMSE (dB)} & \textbf{MAE (dB)} & \textbf{R-Squared} & \textbf{Computation Time (seconds)} \\ \hline\hline
3GPP TR 38.901 Model & 17.25 & 15.93 & -0.81 & 3.5 \\ \hline
ITU-R IMT-2020 Model & 11.13 & 10.51 & 0.25 & 3.4 \\ \hline
Ray-Tracing with Sionna & 26.96 & 25.74 & -3.42 & \(>\)\ 690 \\ \hline
Kriging & 12.02 & 13.65 & 0.15 & 39 \\ \hline
FCNN & 10.59 & 10.23 & 0.24 & 39.4 \\ \hline
PINN with 3GPP TR 38.901 Model & 23.80 & 17.69 & -3.02 & 9.5 (With Early Stopping)\\ \hline
PINN with ITU-R IMT-2020 Model & 12.35& 13.35 & 0.21 & 9.9 (With Early Stopping)\\ \hline
ReVeal & 1.95 & 2.15 & 0.95 & 8.9 (With Early Stopping)\\ \hline
\end{tabular}
\end{table*}

\section{Experimental Results}
\label{sec:results}

The computational experiments for evaluating ReVeal were conducted on a workstation equipped with an Intel® Xeon® processor operating at 3.40\,GHz and has 32\,GB DDR4 RAM. ReVeal is compared to representative 
stochastic models such as the 
3GPP TR 38.901~\cite{3gpp_ts_38.901} and ITU-R IMT-2020~\cite{etsi_tr_138901}. Deterministic models such as ray tracing using Sionna~\cite{sionna} has also been implemented for the part of ARA where we have collected real-world measurement data. 
Additionally, ReVeal has been benchmarked against classical machine learning models commonly used for generating REMs, including the Kriging and standard Fully Connected Neural Network (FCNN) algorithms. 
To characterize the importance of using the PDE form of the path loss model in ReVeal, we also compare ReVeal with its variants where the physics models include 3GPP TR 38.901 and ITU-R IMT-2020, and we denote these models as PINN with 3GPP TR 38.901 Model and PINN with ITU-R IMT-2020 Model, respectively. 
Unless mentioned otherwise, all the PINNs (ReVeal included) use a $\lambda$ value of 0.9. 

Table~\ref{tab: metrics} shows the performance of different benchmark algorithms used for channel modeling. The table includes several common metrics such as root mean squared error (RMSE), mean absolute error (MAE), R-squared, and computation time for each model. 
ReVeal achieves an RMSE of the order of magnitude smaller than other methods, and is accomplished with relatively small computation time. 
    While the statistical models of 3GPP TR 38.901 and ITU-R IMT-2020 have relatively low computation times of 3.5\,s and 3.4\,s, respectively, and have large modeling errors, 
    with an RMSE of 17.25\,dB and 11.13\,dB, respectively. The low computation complexity and high prediction errors in statistical models are due to the simplicity of using predefined formulas capturing the statistical signal propagation behavior and without capturing the specific environment features (e.g., vegetation, trees, and buildings) in ARA at the time of evaluation. 
The computation time for data-driven models such as Kriging and FCNN is higher than that of statistical models, however, still lower than the deterministic ray-tracing model. The ray-tracing model is the most computationally expensive one since it needs to model the environment details and trace individual path between the transmitter and the receiver. 

\begin{figure}[!htbp]
        \centering
        \includegraphics[width=1\columnwidth]{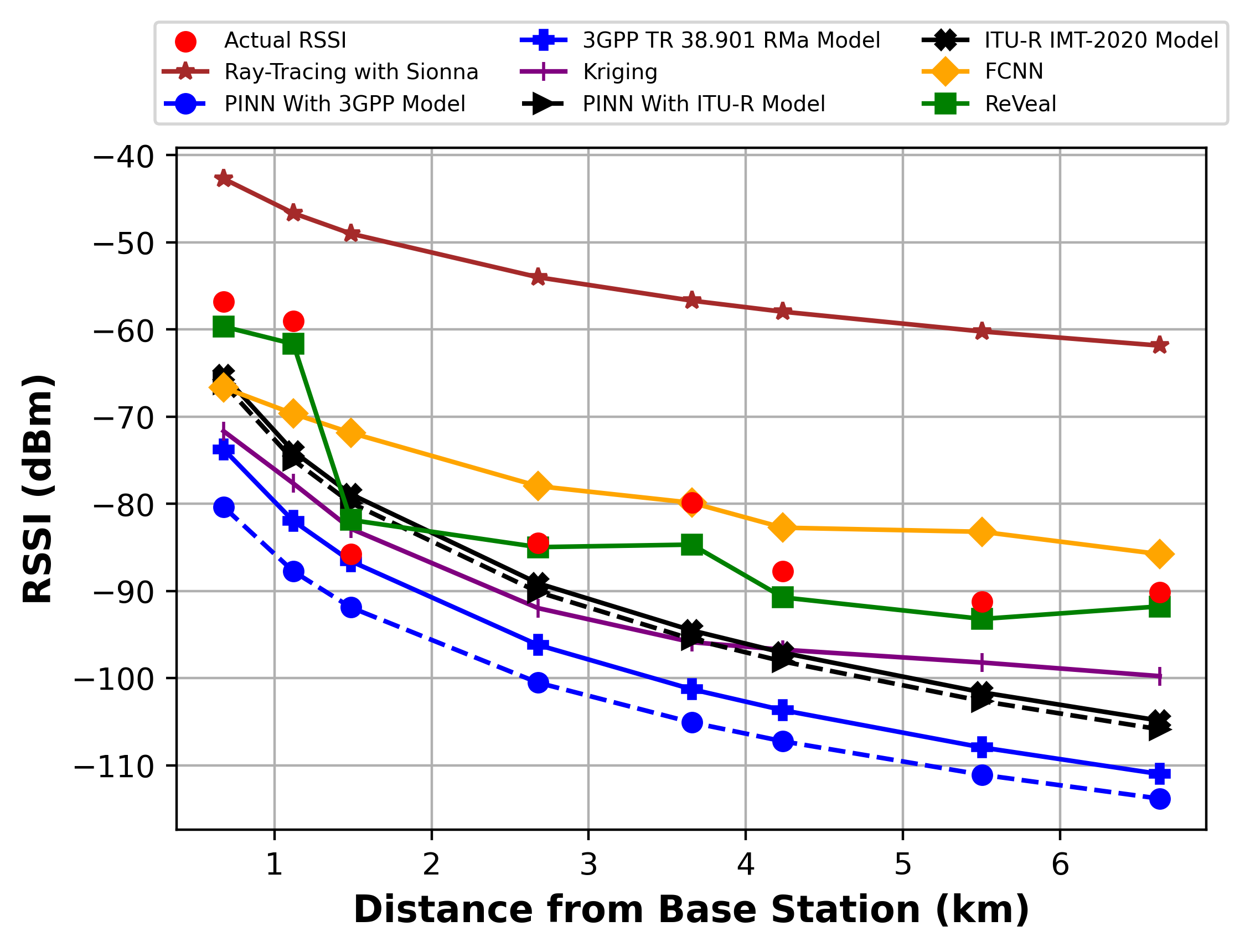}
        \caption{Comparison of predicted and actual RSSI values at different distances from the BS 
        }
        \label{fig: RSSI_comparison}
\end{figure}

For locations at varying distances from the Skylark base station~(BS), Figure~\ref{fig: RSSI_comparison} 
plots the actual RSSI and the RSSI values predicted by different methods. 
    It can be seen that 
    ReVeal 
    performs well at almost all the distances, capturing the variations in RSSI patterns caused due to shadowing. In particular, ReVeal closely follows the changes in RSSI caused by the terrain conditions, demonstrating its ability to model the real-world signal behavior more accurately than existing methods. 
The traditional statistical 
    models tend to underestimate the signal strength by tens of dBs. Kriging model tends to exhibit higher errors in certain scenarios due to its method of estimating values at unmeasured locations by calculating a weighted average of surrounding spatial points. 
    This interpolation approach assumes that the underlying spatial field follows a specific correlation structure, and any deviation from this assumption, such as in complex environments with significant variations in signal strength, can lead to larger prediction errors. 
Similarly, ray-tracing relies heavily on the precise characterization of propagation scenarios, including factors such as vegetation and building materials. However, obtaining such detailed information in a precise manner for large, complex outdoor settings 
is practically infeasible in real-world settings. As a result, ray-tracing performs badly 
due to the lack of accurate environmental data.
    The PINNs with the 3GPP TR 38.901 Model and ITU-R IMT-2020 Model try to align with the behavior of the underlying statistical models, and 
    the inherent limitations of the underlying models restrict the accuracy of the PINNs in such cases.

\begin{figure}[!htbp]
        \centering
        \includegraphics[width=1\columnwidth]{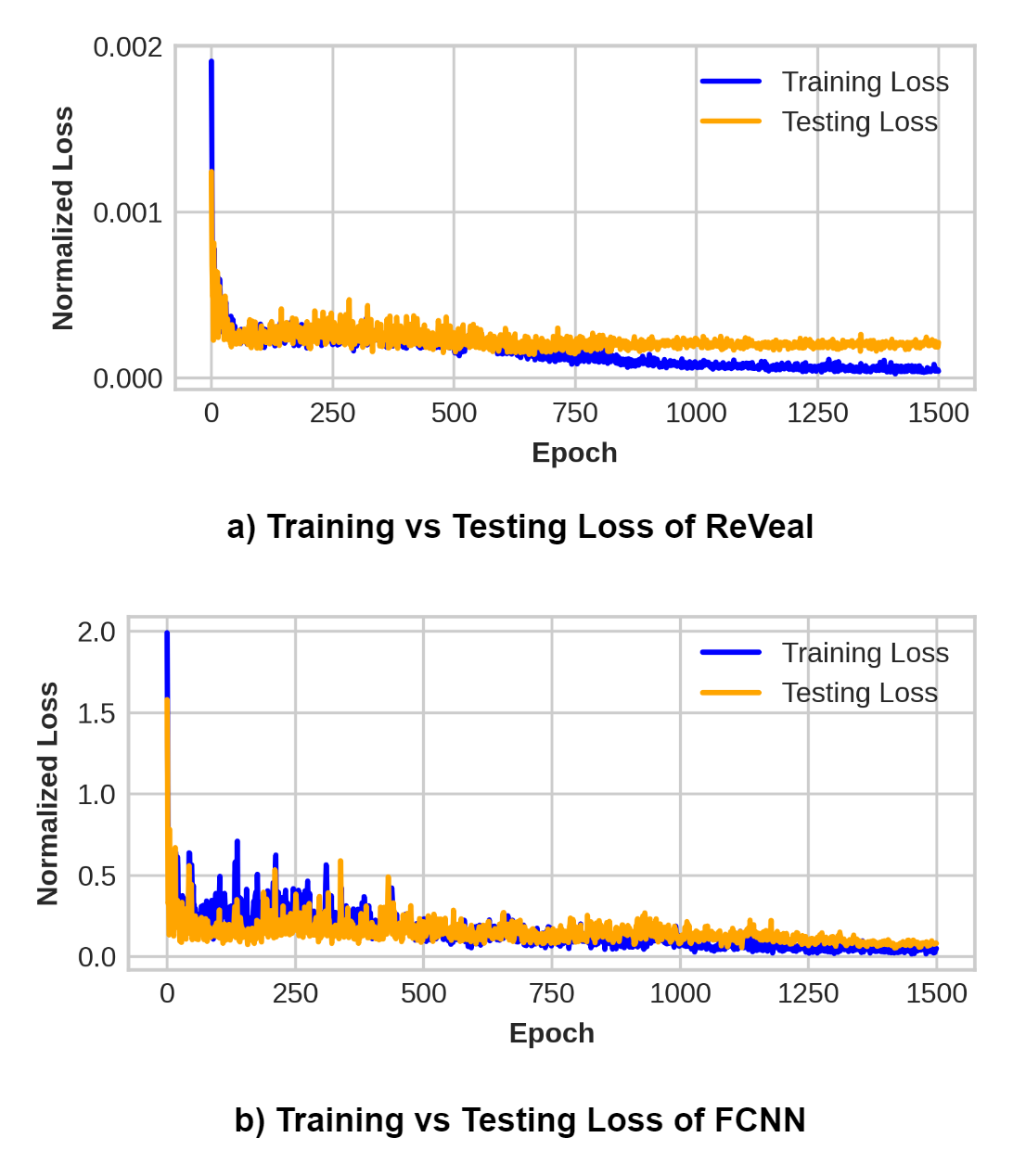}
        \caption{Comparison of Training and Test Loss for ReVeal and FCNN Model 
        }
        \label{fig: Loss_coparison}
\end{figure}

To further characterize the importance of incorporating physical-domain knowledge in radio environment mapping, Figure~\ref{fig: Loss_coparison} 
shows the comparison of ReVeal and FCNN in terms of their training and testing losses. 
In Figure~\ref{fig: Loss_coparison}(a), ReVeal demonstrates rapid convergence, with the training loss dropping quickly and stabilizing close to 300 epochs. The training and testing losses in ReVeal start at 0.0020, whereas it is more than 2 in FCNN. This shows that physics-driven loss helps the neural network learn in a better way and improve the overall performance of the neural network in ReVeal. 
    Figure~\ref{fig: Loss_coparison}(b) also shows that the FCNN model experiences more fluctuations and has higher errors, with both the training and testing losses decreasing more slowly and erratically, suggesting that FCNN struggles to converge and generalize as effectively as ReVeal.

\begin{figure}[!htbp]
        \centering
        \includegraphics[width=1\columnwidth]{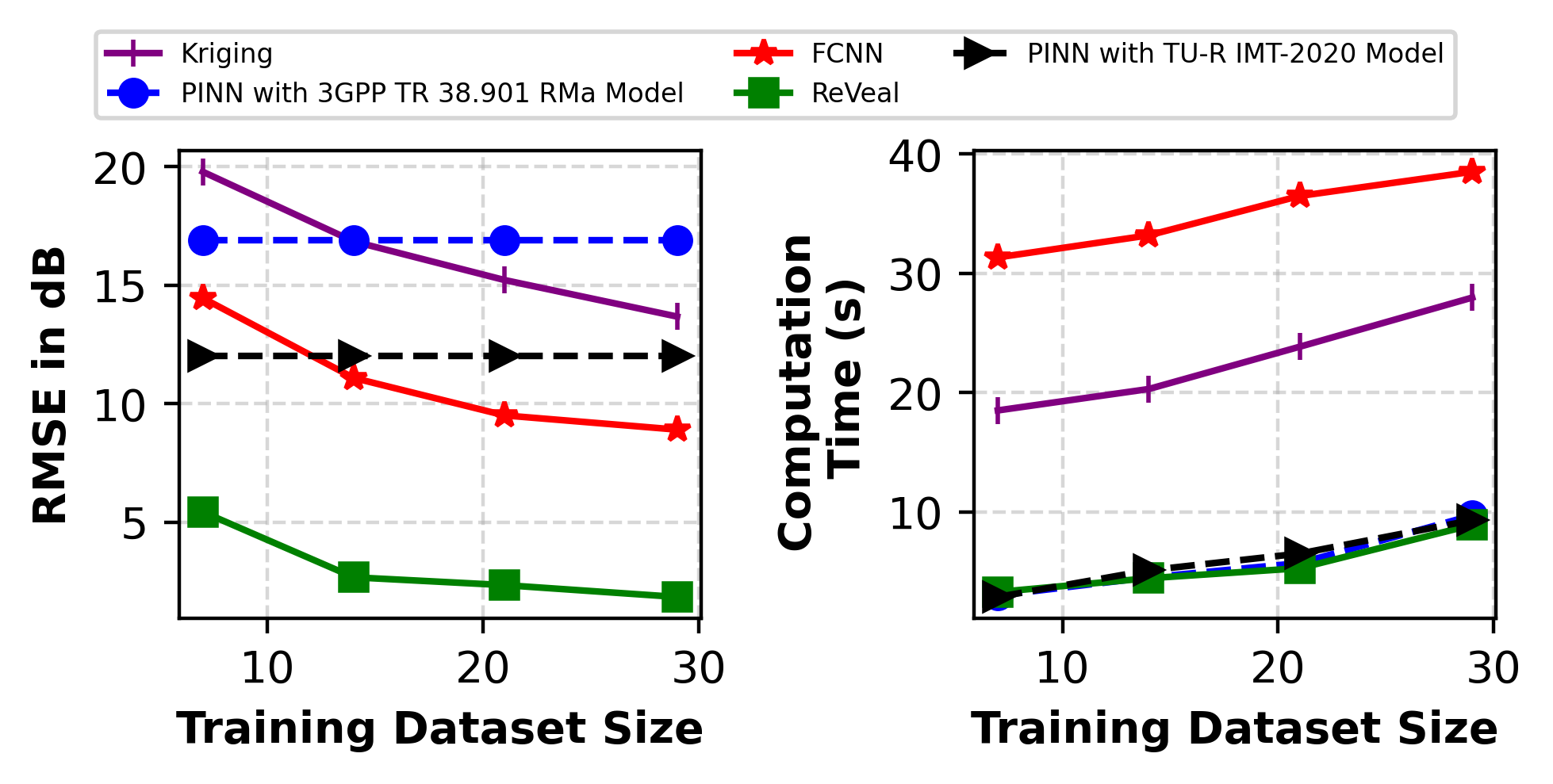}
        \caption{Performance Analysis with other Spectrum Cartography Techniques
        }
        \label{fig: Performance_Analysis}
\end{figure}

To characterize ReVeal's capability of generating accurate REMs with sparse training samples, Figure \ref{fig: Performance_Analysis} 
shows the performance of different data-driven models as sample size varies, where the the PINN models (ReVeal included) use a $\lambda$ value of 0.99.
We see that the RMSE decreases as we increase the number of training samples. 
The PINN models that utilize statistical models as their underlying physics framework show no improvement, as they tend to replicate the behavior of the statistical model itself. With 16 spatial samples, the average RMSE observed while using ReVeal was slightly over 5\,dB, while the RMSE reduced to 1.3\,dB when 30~training samples were used. 
As expected, the computation complexity increases as we increase the number of sample points. 
With early stopping where we stop the training once a desired accuracy is achieved, ReVeal could be trained and then used to visualize the radio environment map in around 8.9~seconds. Other methods tend to require more computation time, for instance, with Kriging and FCNN requiring about 28 and 39 seconds, respectively. 

\begin{figure}[!htbp]
        \centering
        \includegraphics[width=1\columnwidth]{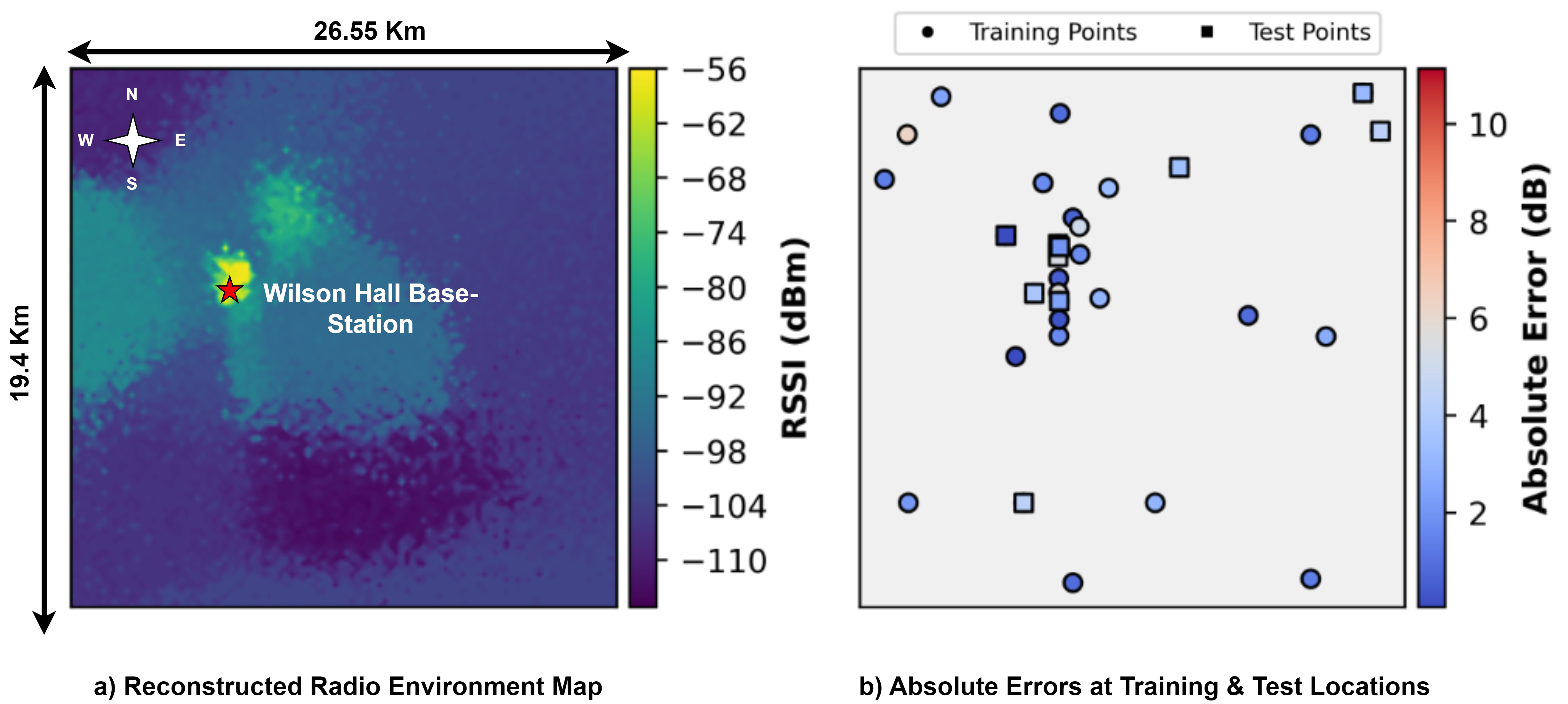}
        \caption{Comparison of Test and Training Error over Regenerated Field 
        }
        \label{fig: Error_comparison}
\end{figure}

\begin{figure}[!htbp]
        \centering
        \includegraphics[width=0.8\columnwidth]{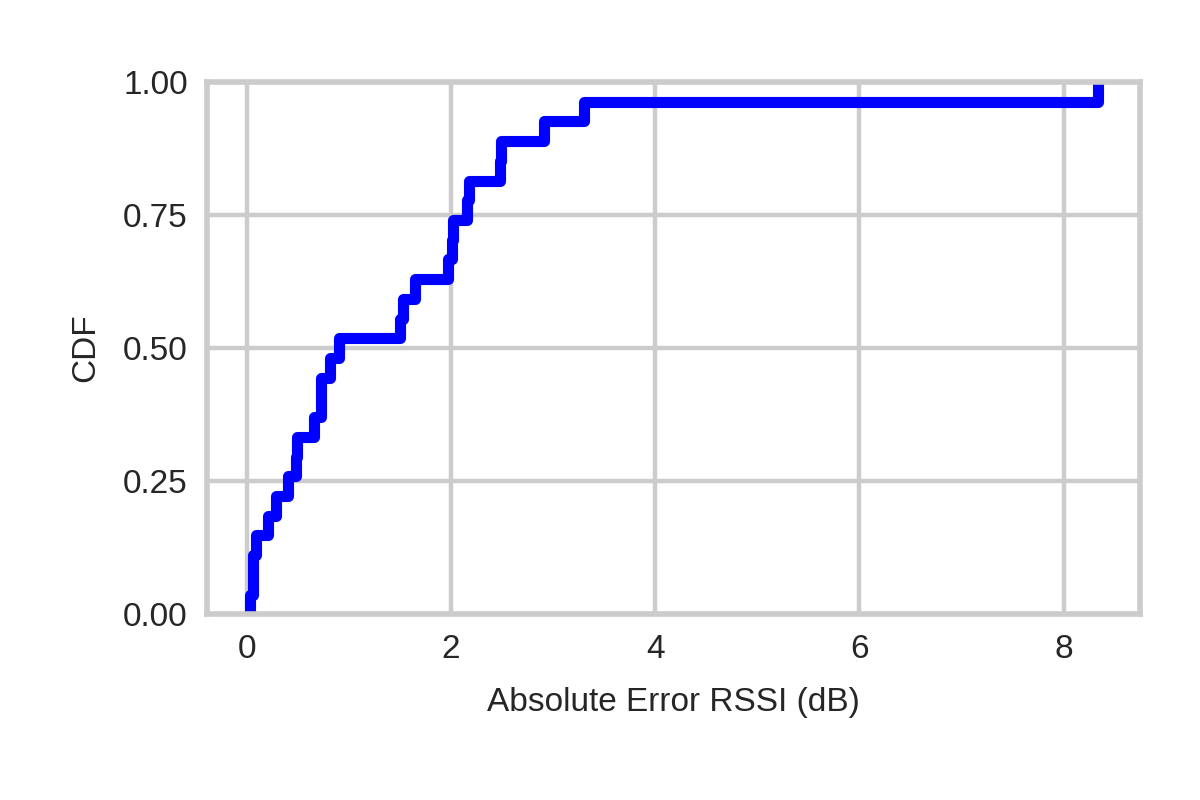}
        \caption{Empirical CDF of Absolute Errors 
        }
        \label{fig: CDF}
\end{figure}

To gain insight into the modeling accuracy of ReVeal, Figure~\ref{fig: Error_comparison}-a 
displays the reconstructed RSSI map for a $\lambda$ value of 0.999. We see that the signal strength varies spatially, with a clear high-signal region surrounded by weaker areas. This spatial variability highlights ReVeal's ability to capture signal propagation patterns effectively. 
    Figure~\ref{fig: Error_comparison}-b illustrates the absolute error for both the training points (circles) and testing points (squares), where the color intensity corresponds to the magnitude of the error. Overall, the error remains relatively low across most locations, indicating good generalization performance. 
To illustrate the probability distribution of the absolute modeling error in ReVeal, Figure~\ref{fig: CDF} 
presents its Empirical Cumulative Distribution Function (ECDF). The 25 percentile, median (50 percentile), and 75 percentile are 1.02\,dB, 1.31\,dB, and 2.39\,dB, respectively.

\begin{figure}[!htbp]
        \centering
        \includegraphics[width=.8\columnwidth]{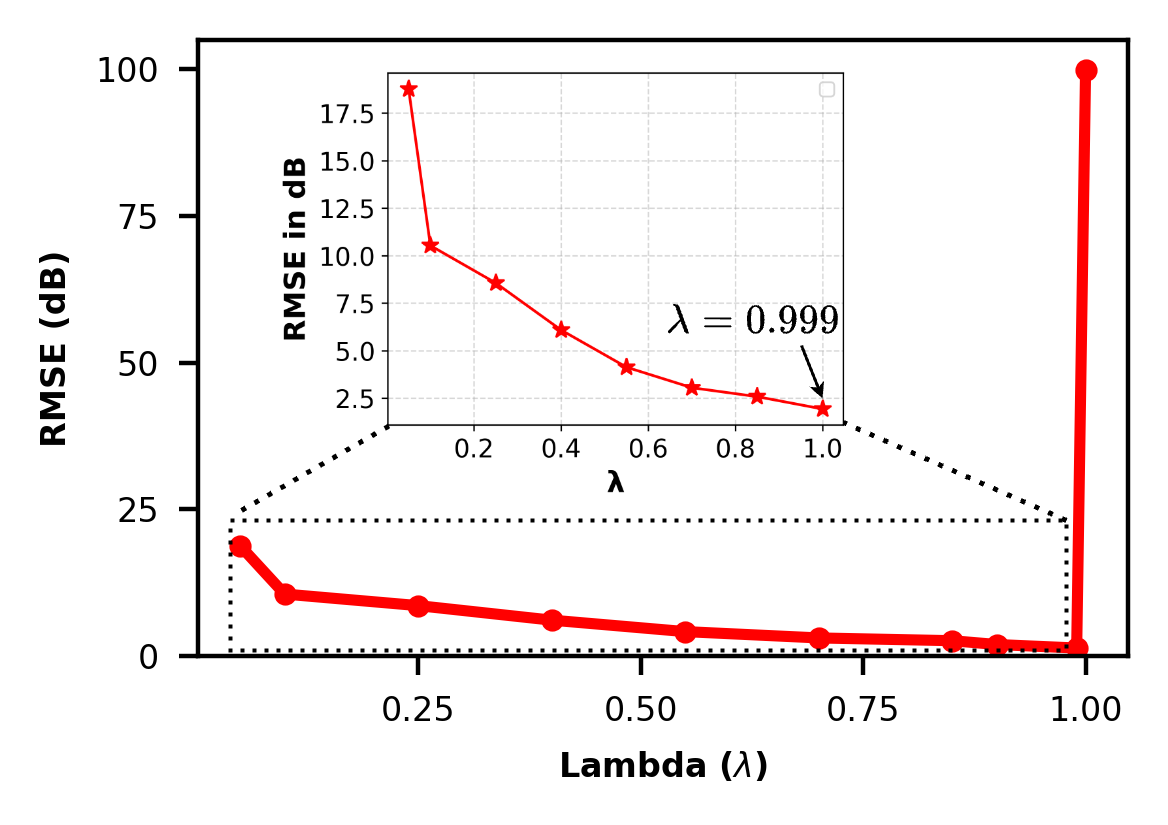}
        \caption{Impact of Varying $\lambda$ on ReVeal Performance }
        \label{fig: lambda_comparison}
\end{figure}

Figure~\ref{fig: lambda_comparison} 
shows the impact of change in \(\lambda\) in Equation~\eqref{eq:total} on the overall RMSE. The parameter \(\lambda\) controls the trade-off between the data-driven loss and the physics-driven loss in the overall loss function. As \(\lambda\) increases, more weight is given to the physics-driven loss compared to the data-driven loss, and hence, the overall error starts to reduce in general. 
    However, when \(\lambda\) becomes 1, data-driven loss ($L_d$) will be completely disregarded, and the RMSE becomes very large; this highlights the critical role of ($L_d$) in guiding the model towards a high-performance 
    solution. In particular, the data-driven loss term ensures that the model adheres to the real-world measurements while maintaining consistency with the governing-physics PDE.
The aforementioned behavior highlights the importance of fine-tuning $\lambda$ to effectively balance the contributions of data-driven and physics-driven loss terms in PINNs for accurate prediction.

\section{Concluding Remarks}
\label{sec:conclusion}

Radio environment mapping 
from a sparse set of spatial measurements plays an important 
role in dynamic spectrum sharing. In this study, 
we have derived a partial-differential-equation (PDE) form of the wireless path loss model and then used the PDE model to design ReVeal, a physics-informed neural network for accurate radio environment mapping in real-world settings. 
    By integrating domain-specific physical principles into the neural network architecture, ReVeal addressed the challenge of sparse measurements and achieved high accuracy and efficiency. 
ReVeal has been evaluated using the first-of-its-kind ARA wireless testbed, demonstrating significant improvements over existing methods as well as its potential for real-world applications. 
    ReVeal enables interesting future research such as applying ReVeal to the design of dynamic spectrum sharing mechanisms and incorporating temporal dynamics to ReVeal for spectrum usage trend prediction. 


\section*{Acknowledgment}
{This work is supported in part by the NSF awards 2130889,  2112606, 2212573, 2229654, and 2232461, NIFA award 202167021-33775, and PAWR Industry Consortium. Consortium. The authors would also like to thank the Microsoft Research team, particularly Ranveer Chandra, Tusher Chakraborty, and Suraj Jog, for intriguing discussions on related topics. 
}

\bibliographystyle{IEEEtran}
\bibliography{references}

\begin{thebibliography}{10}
\providecommand{\url}[1]{#1}
\csname url@samestyle\endcsname
\providecommand{\newblock}{\relax}
\providecommand{\bibinfo}[2]{#2}
\providecommand{\BIBentrySTDinterwordspacing}{\spaceskip=0pt\relax}
\providecommand{\BIBentryALTinterwordstretchfactor}{4}
\providecommand{\BIBentryALTinterwordspacing}{\spaceskip=\fontdimen2\font plus
\BIBentryALTinterwordstretchfactor\fontdimen3\font minus \fontdimen4\font\relax}
\providecommand{\BIBforeignlanguage}[2]{{%
\expandafter\ifx\csname l@#1\endcsname\relax
\typeout{** WARNING: IEEEtran.bst: No hyphenation pattern has been}%
\typeout{** loaded for the language `#1'. Using the pattern for}%
\typeout{** the default language instead.}%
\else
\language=\csname l@#1\endcsname
\fi
#2}}
\providecommand{\BIBdecl}{\relax}
\BIBdecl

\bibitem{Magazine}
M.~S. et. al., ``Wireless spectrum in rural farmlands: Status, challenges and opportunities,'' \emph{arXiv preprint}, vol. 2407.04561, 2024, [Online]. Available: \url{https://arxiv.org/abs/2407.04561}.

\bibitem{DB_Critique}
\BIBentryALTinterwordspacing
R.~Ramjee, S.~Roy, and K.~Chintalapudi, ``A critique of {FCC}'s {TV} white space regulations,'' vol.~20, no.~1, 2016. [Online]. Available: \url{https://doi.org/10.1145/2972413.2972421}
\BIBentrySTDinterwordspacing

\bibitem{DB_Implementation}
M.~Ante, J.~Molina, E.~Trinidad, and L.~Materum, ``A survey and comparison of {TV} white space implementations in {Japan, the Philippines, Singapore, the united kingdom, and the United States},'' \emph{International Journal of Advanced Technology and Engineering Exploration}, vol.~8, pp. 780--796, 07 2021.

\bibitem{Bhattarai2018}
\BIBentryALTinterwordspacing
S.~Bhattarai, ``Spectrum efficiency and security in dynamic spectrum sharing,'' Ph.D. dissertation, Virginia Tech, Apr. 2018, doctoral Dissertation. [Online]. Available: \url{http://hdl.handle.net/10919/82872}
\BIBentrySTDinterwordspacing

\bibitem{vtechworks_2024}
\BIBentryALTinterwordspacing
I.~for Critical~Technology and V.~T. Applied~Science, ``Policy framework for spectrum sharing in the {3.1--3.55 GHz} band,'' 2024, accessed: 2024-11-15. [Online]. Available: \url{https://vtechworks.lib.vt.edu/items/59202f41-6b55-4247-ad3e-4406804f7d60}
\BIBentrySTDinterwordspacing

\bibitem{Spectrum_carography}
\BIBentryALTinterwordspacing
Y.~S. Reddy, A.~Kumar, O.~J. Pandey, and L.~R. Cenkeramaddi, ``Spectrum cartography techniques, challenges, opportunities, and applications: A survey,'' \emph{Pervasive Mob. Comput.}, vol.~79, no.~C, Jan. 2022. [Online]. Available: \url{https://doi.org/10.1016/j.pmcj.2021.101511}
\BIBentrySTDinterwordspacing

\bibitem{cartography_techniques}
\BIBentryALTinterwordspacing
Y.~Teganya, D.~Romero, L.~M.~L. Ramos, and B.~Beferull-Lozano, ``Location-free spectrum cartography,'' \emph{Trans. Sig. Proc.}, vol.~67, no.~15, p. 4013–4026, Aug. 2019. [Online]. Available: \url{https://doi.org/10.1109/TSP.2019.2923151}
\BIBentrySTDinterwordspacing

\bibitem{Subash}
\BIBentryALTinterwordspacing
S.~Timilsina, S.~Shrestha, and X.~Fu, ``Quantized radio map estimation using tensor and deep generative models,'' \emph{Trans. Sig. Proc.}, vol.~72, p. 173–189, Nov. 2023. [Online]. Available: \url{https://doi.org/10.1109/TSP.2023.3336179}
\BIBentrySTDinterwordspacing

\bibitem{SC_survey}
D.~Romero and S.-J. Kim, ``Radio map estimation: A data-driven approach to spectrum cartography,'' \emph{IEEE Signal Processing Magazine}, vol.~39, no.~6, pp. 53--72, 2022.

\bibitem{NTIA_spectrum_cartography}
\BIBentryALTinterwordspacing
C.~R. Dietlein, ``Wide-area spectrum cartography,'' National Telecommunications and Information Administration, Institute for Telecommunication Sciences, Boulder, CO, USA, Tech. Rep., accessed: 2024-11-21. [Online]. Available: \url{https://its.ntia.gov/}
\BIBentrySTDinterwordspacing

\bibitem{Block_tensor_decomposition}
H.~Sun and J.~Chen, ``Integrated interpolation and block-term tensor decomposition for spectrum map construction,'' \emph{IEEE Transactions on Signal Processing}, vol.~72, pp. 3896--3911, 2024.

\bibitem{ARA_design_implementation}
\BIBentryALTinterwordspacing
T.~U. Islam, J.~O. Boateng, M.~Nadim, G.~Zu, M.~Shahid, X.~Li, T.~Zhang, S.~Reddy, W.~Xu, A.~Atalar, V.~Lee, Y.-F. Chen, E.~Gosling, E.~Permatasari, C.~Somiah, Z.~Meng, S.~Babu, M.~Soliman, A.~Hussain, D.~Qiao, M.~Zheng, O.~Boyraz, Y.~Guan, A.~Arora, M.~Selim, A.~Ahmad, M.~B. Cohen, M.~Luby, R.~Chandra, J.~Gross, and H.~Zhang, ``Design and implementation of {ARA} wireless living lab for rural broadband and applications,'' 2024. [Online]. Available: \url{https://arxiv.org/abs/2408.00913}
\BIBentrySTDinterwordspacing

\bibitem{Channel_measurement_survey}
\BIBentryALTinterwordspacing
J.~Zhang, J.~Lin, P.~Tang, Y.~Zhang, H.~Xu, T.~Gao, H.~Miao, Z.~Chai, Z.~Zhou, Y.~Li, H.~Gong, Y.~Liu, Z.~Yuan, X.~Liu, L.~Tian, S.~Yang, L.~Xia, G.~Liu, and P.~Zhang, ``Channel measurement, modeling, and simulation for {6G}: A survey and tutorial,'' \emph{arXiv preprint arXiv:2305.16616}, 2023. [Online]. Available: \url{https://arxiv.org/abs/2305.16616}
\BIBentrySTDinterwordspacing

\bibitem{Tataria2020}
\BIBentryALTinterwordspacing
H.~Tataria, K.~Haneda, A.~F. Molisch, M.~Shafi, and F.~Tufvesson, ``Standardization of propagation models for terrestrial cellular systems: A historical perspective,'' \emph{International Journal of Wireless Information Networks}, vol.~27, pp. 340--364, 2020. [Online]. Available: \url{https://link.springer.com/article/10.1007/s10776-020-00500-9}
\BIBentrySTDinterwordspacing

\bibitem{ray_tracing_based_model}
J.-H. Lee and A.~F. Molisch, ``A scalable and generalizable pathloss map prediction,'' \emph{IEEE Transactions on Wireless Communications}, vol.~23, no.~11, pp. 17\,793--17\,806, 2024.

\bibitem{Nueral_Ray_Tracing}
\BIBentryALTinterwordspacing
T.~Orekondy, P.~Kumar, S.~Kadambi, H.~Ye, J.~Soriaga, and A.~Behboodi, ``{WiNeRT}: Towards neural ray tracing for wireless channel modelling and differentiable simulations,'' in \emph{Proceedings of the 11th International Conference on Learning Representations (ICLR 2023)}, 2023. [Online]. Available: \url{https://openreview.net/forum?id=tPKKXeW33YU}
\BIBentrySTDinterwordspacing

\bibitem{RDZ}
M.~Zheleva, C.~R. Anderson, M.~Aksoy, J.~T. Johnson, H.~Affinnih, and C.~G. DePree, ``Radio dynamic zones: Motivations, challenges, and opportunities to catalyze spectrum coexistence,'' \emph{IEEE Communications Magazine}, vol.~61, no.~6, pp. 156--162, 2023.

\bibitem{Kriging}
P.~Maiti and D.~Mitra, ``Complexity reduction of ordinary {Kriging} algorithm for {3D REM} design,'' \emph{Physical Communication}, vol.~55, p. 101912, 10 2022.

\bibitem{DeepREM}
A.~Chaves-Villota and C.~A. Viteri-Mera, ``{DeepREM}: Deep-learning-based radio environment map estimation from sparse measurements,'' \emph{IEEE Access}, vol.~11, pp. 48\,697--48\,714, 2023.

\bibitem{ProSpire}
S.~Sarkar, D.~Guo, and D.~Cabric, ``{ProSpire}: Proactive spatial prediction of radio environment using deep learning,'' in \emph{2023 20th Annual IEEE International Conference on Sensing, Communication, and Networking (SECON)}, 2023, pp. 177--185.

\bibitem{CNN}
H.~Cheng, H.~Lee, and S.~Ma, ``{CNN}-based indoor path loss modeling with reconstruction of input images,'' in \emph{2018 International Conference on Information and Communication Technology Convergence (ICTC)}, 2018, pp. 605--610.

\bibitem{U_NET}
M.~Mallik, S.~Kharbech, T.~Mazloum, S.~Wang, J.~Wiart, D.~P. Gaillot, and L.~Clavier, ``{EME-Net}: A {U-net-based} indoor {EMF} exposure map reconstruction method,'' in \emph{2022 16th European Conference on Antennas and Propagation (EuCAP)}, 2022, pp. 1--5.

\bibitem{GAN}
\BIBentryALTinterwordspacing
X.~Han, L.~Xue, F.~Shao, and Y.~Xu, ``A power spectrum maps estimation algorithm based on generative adversarial networks for underlay cognitive radio networks,'' \emph{Sensors}, vol.~20, no.~1, 2020. [Online]. Available: \url{https://www.mdpi.com/1424-8220/20/1/311}
\BIBentrySTDinterwordspacing

\bibitem{PINN_to_PIKAN}
\BIBentryALTinterwordspacing
J.~D. Toscano, V.~Oommen, A.~J. Varghese, Z.~Zou, N.~A. Daryakenari, C.~Wu, and G.~E. Karniadakis, ``From {PINNs to PIKANs}: Recent advances in physics-informed machine learning,'' \emph{arXiv preprint arXiv:2410.13228}, 2024. [Online]. Available: \url{https://arxiv.org/abs/2410.13228}
\BIBentrySTDinterwordspacing

\bibitem{Possion_FEM}
\BIBentryALTinterwordspacing
D.~E. Johnson, ``Numerical solutions to {Poisson} equations using the finite-difference method,'' \emph{IEEE Antennas and Propagation Magazine}, vol.~56, no.~6, pp. 158--162, 2014. [Online]. Available: \url{https://ieeexplore.ieee.org/document/6931698}
\BIBentrySTDinterwordspacing

\bibitem{PINN_RAISSI}
\BIBentryALTinterwordspacing
M.~Raissi, P.~Perdikaris, and G.~Karniadakis, ``Physics-informed neural networks: A deep learning framework for solving forward and inverse problems involving nonlinear partial differential equations,'' \emph{Journal of Computational Physics}, vol. 378, pp. 686--707, 2019. [Online]. Available: \url{https://www.sciencedirect.com/science/article/pii/S0021999118307125}
\BIBentrySTDinterwordspacing

\bibitem{Autotune}
P.~Koch, O.~Golovidov, S.~Gardner, B.~Wujek, J.~Griffin, and Y.~Xu, ``Autotune: A derivative-free optimization framework for hyperparameter tuning,'' 04 2018.

\bibitem{SMAC}
\BIBentryALTinterwordspacing
F.~Hutter, H.~H. Hoos, and K.~Leyton-Brown, ``Sequential model-based optimization for general algorithm configuration,'' in \emph{Proceedings of the 5th International Conference on Learning and Intelligent Optimization}, ser. LION'05.\hskip 1em plus 0.5em minus 0.4em\relax Berlin, Heidelberg: Springer-Verlag, 2011, p. 507–523. [Online]. Available: \url{https://doi.org/10.1007/978-3-642-25566-3_40}
\BIBentrySTDinterwordspacing

\bibitem{optuna}
T.~Akiba, S.~Sano, T.~Yanase, T.~Ohta, and M.~Koyama, ``{O}ptuna: A next-generation hyperparameter optimization framework,'' in \emph{The 25th ACM SIGKDD International Conference on Knowledge Discovery \& Data Mining}, 2019, pp. 2623--2631.

\bibitem{spatial_sampling}
\BIBentryALTinterwordspacing
A.~Ivanov, K.~Tonchev, V.~Poulkov, A.~Manolova, and A.~Vlahov, ``Limited sampling spatial interpolation evaluation for {3D} radio environment mapping,'' \emph{Sensors}, vol.~23, no.~22, 2023. [Online]. Available: \url{https://www.mdpi.com/1424-8220/23/22/9110}
\BIBentrySTDinterwordspacing

\bibitem{grafstra2012spatially}
A.~Grafstr{\~A}, N.~L. Lundstr{\~A}, L.~Schelin \emph{et~al.}, ``Spatially balanced sampling through the pivotal method,'' 2012.

\bibitem{3gpp_ts_38.901}
\BIBentryALTinterwordspacing
3GPP, ``{3GPP TS 38.901}---{Study} on channel model for frequencies from {0.5 to 100 GHz},'' 2021, accessed: 2024-12-13. [Online]. Available: \url{https://portal.3gpp.org/desktopmodules/Specifications/SpecificationDetails.aspx?specificationId=3173}
\BIBentrySTDinterwordspacing

\bibitem{etsi_tr_138901}
\BIBentryALTinterwordspacing
ETSI, ``{ETSI TR 138 901 V14.0.0}--- technical report: Study on channel model for frequencies from {0.5 to 100 GHz},'' 2020, accessed: 2024-12-13. [Online]. Available: \url{https://www.etsi.org/deliver/etsi_tr/138900_138999/138901/14.00.00_60/tr_138901v140000p.pdf}
\BIBentrySTDinterwordspacing

\bibitem{sionna}
J.~Hoydis, S.~Cammerer, F.~{Ait Aoudia}, A.~Vem, N.~Binder, G.~Marcus, and A.~Keller, ``Sionna: An open-source library for next-generation physical layer research,'' \emph{arXiv preprint}, Mar. 2022.

\end{thebibliography}

\end{document}